\documentclass[aps, twocolumn, prl, 10pt, showkeys, showpacs]{revtex4-1}
\usepackage{amssymb}
\usepackage{amsmath}
\DeclareMathOperator{\tr}{tr}
\begin{document}
\title{Modelling gravity on a hyper-cubic lattice}
\author{Kyle Tate}
\email{kyle.tate@msor.vuw.ac.nz}
\author{Matt Visser}
\email{matt.visser@msor.vuw.ac.nz}
\affiliation{School of Mathematics, Statistics, and Operations Research, \\
Victoria University of Wellington, PO Box 600, Wellington 6140, New Zealand }
\date{6 June 2012; \LaTeX-ed \today}
\begin{abstract}
We present an elegant and simple dynamical model of symmetric, non-degenerate $n \times n$ matrices of fixed signature defined on a $n$-dimensional hyper-cubic lattice with nearest-neighbor interactions. We show how this model is related to General Relativity, and discuss multiple ways in which it can be useful for studying gravity, both classical and quantum. In particular, we show that the dynamics of the model when all matrices are close to the identity corresponds exactly to a finite-difference discretization of weak-field gravity in harmonic  gauge. We also show that the action which defines the full dynamics of the model corresponds to the Einstein--Hilbert action to leading order in the lattice spacing, and use this observation to define a lattice analogue of the Ricci scalar and Einstein tensor. Finally, we perform a mean-field analysis of the statistical mechanics of this model. 
\end{abstract}
\keywords{Lattice models, Gravity}
\pacs{04.60.Nc;  04.50.-h;   04.20.-q; 04.60.-m}
\maketitle
\def\I{{\mathbb{I}}}
\def\widebar{\overline}
\def\R{\mathcal{R}}
\def\G{\mathcal{G}}
\def\O{\mathcal{O}}

\noindent\paragraph{Introduction:}
Lattice models of gravity are typically defined using some discretization which is simultaneously coordinate and background independent. The most popular such discretization is the Regge calculus~\cite{Regge:1961px, Williams:1996jb}, used to study both classical general relativity (GR)~\cite{Sorkin:1975ah, Barrett:1994ks}, as well as to build models of quantum gravity based on dynamical simplices~\cite{Rocek:1982fr, Williams:1986hx, Hamber:1985gw, Gross:1990fq, David:1992jw, Ambjorn:1998xu, Loll:2000my, Ambjorn:2002gr, Ambjorn:2005db, Ambjorn:2005qt, Ambjorn:2006jf, Tate:2011ct, Tate:2011rm}. Another discretization, based on a topological lattice, is~\cite{Loll:1995mg}, (see also~\cite{Loll:1998aj}), however this discretization is designed to preserve some form of diffeomorphism invariance, and so is also both coordinate and background independent. 

In this Letter we shall instead present a discrete model for gravity which is defined on a regular, in fact hyper-cubic, coordinate lattice. The implied background structure may be anathema to GR purists,  however we will argue that this is still a useful thing to do, and can be usefully utilized to study GR. 

That having a preferred background might not be entirely implausible may be inferred from a number of observations: 
(1) Many interesting spacetimes can be put into Kerr--Schild form, which has a natural background~\cite{exact}.  
(2) Many interesting spacetimes can be put into Painleve--Gullstrand~\cite{acoustic} and/or de~Donder (nonlinear harmonic) form~\cite{fock}, both of which possess natural background metrics. 
(3) Many interesting spacetimes can be put into ``relativistic acoustic'' form, based on the ``analogue spacetime'' programme, for which a natural background metric again exists~\cite{carmen}.
(4) Physically interesting black holes can be put into horizon-penetrating coordinates, for which the metric components are finite at the horizon;  the presence of a horizon does not \emph{necessarily} imply ``infinite deviations'' from some assumed background metric \cite{acoustic, petrov}. 
(5) More exotically, recent speculations on ghost-free massive gravitons are most naturally phrased in terms of a combination of foreground and background metric~\cite{massive-graviton}. 
In view of the above, we are willing to at least entertain the notion of background structure, to see how far we can get. 

\noindent\paragraph{Lattice action:} 
Consider a $n$-dimensional hyper-cubic lattice which has defined, at each site $i$,  an $n \times n$ symmetric, non-degenerate matrix ${}^i\!g$, which is physically to be interpreted as the metric. Unless otherwise stated, we will assume in this Letter that the matrix is positive definite, (and hence is a model for a Euclidean-signature Riemannian geometry), however the model can easily be generalized to matrices of any fixed signature. The dynamics of the model is described by a particularly simple and elegant action defined as a sum over nearest neighbor pairs. Let us first define an ``average'' metric linking the sites $i$ and $j$
\begin{equation}
^{ij}\widebar{g} = {  \frac{{}^{j}\!g + {}^i\!g}{2} },
\end{equation}
and then set
\begin{eqnarray}
\label{eq: actiondef}
S &=&  - \xi_n \; \left({s /  L_P}\right)^{n-2}
 \\
&\times& 
 \sum_{\langle ij \rangle}  \left\{ \sqrt{\det\left( ^{ij}\widebar{g}   \right)} - \frac{\sqrt{\det({}^i\!g)}}2 - \frac{\sqrt{\det({}^j\!g)}}2\right\}.
\nonumber
\end{eqnarray}
Here $\langle ij \rangle$ denotes the link joining nearest neighbor sites $i$ and $j$. The lattice spacing is $s$, and the $n$-dimensional Planck length is $L_P$, while $\xi_n$ is some convenient dimension-dependent normalizing constant.
Note that this action has a symmetry under both rigid (global) $SO(n)$ transformations ${}^i\!g \rightarrow O^T \; {}^i\!g \; O$, and under parity.

\noindent\paragraph{Weak field:} 

Consider small fluctuations
\begin{equation}
\label{eq: linearized}
{}^i\!g = \I + {}^i\!h; \qquad |{}^i\!h|\ll 1.
\end{equation}
To quadratic order in $h$ we have 
\begin{eqnarray}
\label{eq: neighbourexp}
\sqrt{\det\left( ^{ij}\widebar{g}  \right)} &=& 1 + \frac14 \, \tr\left[ {}^i\!h + {}^j\!h \right]+ \frac{1}{32} \,\tr\left[ {}^i\!h + {}^j\!h \right]^2 \nonumber \\
&&- \frac{1}{16} \, \tr\left[( {}^i\!h + {}^j\!h)^2 \right] + \O(h^3),
\end{eqnarray}
and
\begin{equation}
\label{eq: indexp}
\frac12 \, \sqrt{\det({}^i\!g) } = \frac12 + \frac14 \tr\left[{}^i\!h \right] + \frac{1}{16} \,  \tr\left[{}^i\!h \right]^2 - \frac18  \tr\left[{}^i\!h^2 \right] + \O(h^3).
\end{equation}
Thus to quadratic order in $h$ we have
\begin{equation}
\label{eq: linact}
S \propto
\sum_{i} \, \sum_{j : \langle ij \rangle} \left( \tr\left[ \, ({}^j\!h - {}^i\!h)^2 \right] - \frac12 \tr\left[ \,{}^j\!h - {}^i\!h \right]^2 \right) +\O(h^3).
\end{equation}
One might worry that the matrices were only expanded to linear order; however repeating the calculation and keeping the quadratic terms gives the same result. Taking the lattice spacing to zero the finite differences become derivatives, and the action (up to an arbitrary multiplicative constant) is given by
\begin{equation}
\label{eq: contact}
S \propto \int_{\mathbb{R}^n} d^nx \; \left( \partial_{\sigma} h_{\mu \nu} \; \partial^{\sigma} h^{\mu \nu} - \frac12 \, \partial^{\mu} h^{\nu}{}_{\nu} \; \partial_{\mu} h^{\sigma}{}_{\sigma} \right) +\O(h^3).
\end{equation}
This is precisely the action for linearized GR in (linearized) harmonic gauge 
\begin{equation}
\label{eq: harmoniclinear}
\partial^{\nu} h_{\mu \nu} - \frac12 \,\partial_{\mu} h^{\sigma}{}_{\sigma} = 0. 
\end{equation}
Thus we see that the lattice action (\ref{eq: actiondef})  can be used to model weak field GR provided the very commonly used  (linearized) harmonic coordinates are adopted. 

\noindent\paragraph{Continuum limit:} 
Naively,  it might seem just a lucky coincidence that the coefficients in the linearized expansion are precisely those needed to recover the GR terms of weak-field gravity. However, we will show that the discrete  lattice action (\ref{eq: actiondef}) has a continuum limit, which is given to leading order by the Einstein--Hilbert action.
 Let us use permutation symmetry to first rewrite the action as
\begin{equation}
\label{eq: actrewrite}
S \propto \sum_{\langle ij \rangle} \left\{ \sqrt{\det\left( ^{ij}\widebar{g} \right)} - \sqrt{\det({}^i\!g) }\right\}.
\end{equation}
We can explicitly factor out a $\sqrt{\det({}^i\!g) }$ to obtain
\begin{equation}
\label{eq: lattEHform}
S = \sum_{i} \sqrt{\det({}^i\!g) } \; \;  {}^i\!S.
\end{equation}
Here we define the site-specific contribution to the action for site $i$ in terms of a sum over  its nearest neighbours 
\begin{equation}
\label{eq: sitecontribution}
{}^i\!S \propto \sum_{j : \langle ij \rangle}  \left(2^{-n/2} \, \sqrt{ \det \left( \I + [{}^i\!g]^{-1} \; [{}^j g] \right)} - 1 \right).
\end{equation}
Let us now work on a continuum manifold with metric $g_{\mu\nu}(x)$ and choose a Riemann normal coordinate system --- such that site $i$ is taken to be the origin, and the metric at site $i$ is ${}^i\!g_{\mu \nu} = \delta_{\mu \nu}$. Let $\ell^\mu_{ij}$ denote the unit vector pointing from site $i$ to site $j$. The coordinate system is such that the geodesics generated by these coordinate unit vectors are straight lines:
\begin{equation}
\label{eq: RNCgeo}
{}^{ij}x^{\mu}(\lambda) = \ell^{\mu}_{ij} \; \lambda.
\end{equation}
Then in the immediate neighborhood of the origin we can construct the vertices of the hyper-cubic lattice such that the nearest neighbors are connected by geodesics of length $s$ and have coordinate locations  ${}^j\!x^{\mu} = \ell^{\mu}_{ij}\; s$. A standard result for Riemann normal coordinates is that to quadratic order
\begin{equation}
\label{eq: latticemetriccont}
{}^{j}\!g_{\mu \nu} = g_{\mu \nu}({}^{ij}x) = \delta_{\mu \nu} - \frac13 \; \R_{\mu \alpha \nu \beta} \, \ell^{\alpha}_{ij} \, \ell^{\beta}_{ij} \, s^2 +\O(s^3), 
\end{equation}
But then
\begin{equation}
\{[{}^i\!g]^{-1} \; [{}^j g]\}^\mu{}_\nu = \delta^\mu{}_\nu - \frac13 \; \R^\mu{}_{\alpha \nu \beta} \, \ell^{\alpha}_{ij} \, \ell^{\beta}_{ij} \, s^2 +\O(s^3), 
\end{equation}
so
\begin{equation}
2^{-n/2} \sqrt{ \det \left( \I + [{}^i\!g]^{-1} \; [{}^j g] \right)}  =  1 - \frac1{12} \; \R_{\alpha \beta} \, \ell^{\alpha}_{ij} \, \ell^{\beta}_{ij} \, s^2 +\O(s^3).
\end{equation}
Hence the site-specific contribution to the action is
\begin{equation}
\label{eq: lattR}
{}^i\!S \propto \sum_{j : \langle ij \rangle}   \R_{\alpha \beta} \; \ell^{\alpha}_{ij} \, \ell^{\beta}_{ij} \, s^2 +\O(s^3).
\end{equation}
Using the easily deduced result
\begin{equation}
\label{eq: sumidentities}
\sum_{j: \langle ij \rangle} \ell^{\alpha}_{ij} \, \ell^{\beta}_{ij} = 2 \, \delta^{\alpha \beta},
\end{equation}
we see that to quadratic order in the lattice spacing (which is also the geodesic distance) the continuum analogue of equation \eqref{eq: sitecontribution} is given by
\begin{equation}
\label{eq: continuumactioncont}
{}^i\!S \propto \R \; s^2 +\O(s^3).
\end{equation}
Thus we see that our lattice action corresponds to the Einstein--Hilbert action,  to leading order in the lattice spacing. 

\paragraph{Lattice Ricci scalar:} 
Choosing a suitable normalization we define
\begin{equation}
\label{eq: lattR}
{}^i\!R = - 6 \sum_{j : \langle ij \rangle}  \left(2^{-n/2} \, \sqrt{ \det \left( \I + [{}^i\!g]^{-1} \; [{}^j g] \right)} - 1 \right).
\end{equation}
Then trivially adapting the discussion above
\begin{equation}
^i\!R = \; \R \; s^2 +\O(s^3). 
\end{equation}
We see that to lowest nontrivial order in the lattice spacing, the discrete quantity ${}^i\!R$ exactly matches its continuum analogue $\R$. 

\paragraph{Strong field EOM:}
The discrete version of the Einstein tensor is easily obtained by computing
\begin{equation}
^i\!G \propto  {1\over\sqrt{\det(^i\!g)}} \; {\delta S\over \delta[^i\!g]}.
\end{equation}
We find
\begin{eqnarray}
\frac{\delta S}{\delta[^i\! g]} = \frac12 \, \sum_{j : \langle ij \rangle} \Bigg( \sqrt{\det \left( ^{ij}\widebar{g} \right)}\, \left( ^{ij}\widebar{g} \right)^{-1}
-  \sqrt{ \det ({}^i\!g)} \, ({}^i\!g)^{-1} \, \Bigg).
\nonumber\\
\end{eqnarray}
After picking a suitable normalization, we set
\begin{equation}
^i\!G =  \, - 6 \sum_{j : \langle ij \rangle} \Bigg[ \sqrt{\det \left( ^{ij}\widebar{g} \right)\over\det(^i\!g)}\, \left( ^{ij}\widebar{g} \right)^{-1}
-  ({}^i\!g)^{-1} \, \Bigg].
\end{equation}
But now (again adopting Riemann normal coordinates at the site $i$, so $^i\!g\to\I$) we have already seen (to quadratic order) the equivalent of
\begin{equation}
\left[^{ij}\widebar{g} \right]^{-1}_{\mu\nu}=  \delta_{\mu\nu} + \frac16  \;  \R_{\mu \alpha \nu \beta} \; \ell^{\alpha}_{ij} \, \ell^{\beta}_{ij} \; s^2 +\O(s^3), 
\end{equation}
and
\begin{equation}
\det\left[ ^{ij}\widebar{g} \right]=  1 - \frac16  \;  \R_{\alpha \beta} \; \ell^{\alpha}_{ij} \, \ell^{\beta}_{ij} \; s^2 +\O(s^3), 
\end{equation}
whence, to quadratic order
\begin{equation}
^i\!G_{\mu\nu} =  \, \frac{1}{2} \sum_{j : \langle ij \rangle} \Bigg[  (\R_{\mu \alpha \nu \beta}  - {1\over2}  \delta_{\mu\nu} \; \R_{\alpha \beta} ) \; \ell^{\alpha}_{ij} \, \ell^{\beta}_{ij} s^2 \Bigg] +\O(s^3).
\end{equation}
So, summing over the nearest neighbour sites $j$, the discrete Einstein tensor is related to the continuum Einstein tensor by
\begin{equation}
^i\!G_{\mu\nu} =  \,   \Bigg[ \R_{\mu\nu}  - {1\over2}  \delta_{\mu\nu} \; \R  \Bigg] s^2 +\O(s^3) = \G_{\mu\nu}\; s^2 +\O(s^3).
\end{equation}

\paragraph{External stress-energy tensor:}
As is usual in lattice models, we can also add an external current $^i\!J$ to probe the dynamics. The most natural object to add is
\begin{equation}
S_J = \sum_i \sqrt{ \det ({}^i\!g)}\; \tr[ ^i\! g \; ^i\!J],
\end{equation}
The external current $^i\!J$ is interpretable in terms of the discrete stress-energy tensor $^iT$ via
\begin{equation}
^iT =  {1\over\sqrt{\det(^i\!g)}} \; {\delta S_J\over \delta[^i\!g]} = {}^i\!J + {1\over2} \;   [ ^i\! g ]^{-1}\; \tr[ ^i\! g \; ^i\!J].
\end{equation}
The strong field discrete Einstein equations in the presence of external stress-energy are then quite simply
\begin{equation}
^i\!G_{\mu\nu} \propto {} ^iT_{\mu\nu}.
\end{equation}
So formally at least, the discrete lattice model contains all the correct ingredients for adequately dealing with large swathes of standard GR.  This procedure clearly generalizes to placing some matter model on the lattice. 

\paragraph{Mean-field analysis:}
In addition to using this lattice model to study classical GR, it can also be used as a discrete model for studying quantum gravity. A first step in this direction is to perform a mean-field analysis of the action \eqref{eq: actiondef} which we now rewrite in terms of the site-specific form as
\begin{eqnarray}
\label{eq: actionrewrite2}
S &=& - \alpha_n\, \sum_{i} \sqrt{\det({}^i\!g) } \\
&& \times \sum_{j : \langle ij \rangle}  \left(2^{-n/2} \, \sqrt{ \det \left( \I + [{}^i\!g]^{-1} \; [{}^j g] \right)} - 1 \right), \nonumber
\end{eqnarray}
where $\alpha_n = \zeta_n (s/L_P)^{n-2}$ is the constant appearing in equation \eqref{eq: actiondef}. This action is translationally invariant and thus we take the mean-field ansatz ---  assuming the physics is dominated by some translation invariant average $M = \langle g \rangle$,  plus small fluctuations 
\begin{equation}
\label{eq: meanfieldansatz}
{}^i\!g = M + \delta [^i\!g].
\end{equation}
This allows us to replace $^j\!g$ with $M$ in the coupling term of the action.  We find
\begin{eqnarray}
\label{eq: fluctexpand}
&&\sum_{j : \langle ij \rangle} \left(2^{-n/2} \, \sqrt{ \det \left( \I + [{}^i\!g]^{-1} \; [{}^j g] \right)} - 1 \right)
\nonumber \\
&& \qquad 
 \to  2n \left(2^{-n/2} \, \sqrt{ \det \left( \I + [{}^i\!g]^{-1} \;M \right)} - 1 \right).
\end{eqnarray}
The total action then becomes
\begin{eqnarray}
\label{eq: mfaction}
S_\mathrm{mf} &=& - \alpha_n \, 2n \, \sum_{i} \sqrt{\det({}^i\!g) }  
\\
&& 
\times \left(2^{-n/2} \, \sqrt{ \det \left( \I + [{}^i\!g]^{-1} \;M \right)} - 1 \right). 
\nonumber
\end{eqnarray}
Thus the mean field partition function
\begin{equation}
\label{eq: mfpartition}
Z_\mathrm{mf} = \int \prod_i d^i\!g \; e^{- \beta S_\mathrm{mf}},
\end{equation}
is given by
\begin{eqnarray}
\label{eq: mfpartitionbefore}
Z_\mathrm{mf} &=& \left( \int dg \; e^{\gamma \, \sqrt{\det g} \left( 2^{-n/2} \, \sqrt{ \det \left( \I + g^{-1} \,M \right)} - 1  \right)} \right)^N\!\!\!,\qquad
\end{eqnarray}
where $\gamma = 2n \, \alpha_n \, \beta$. Now performing a change variables $g = \sqrt{M} \, \tilde{g} \sqrt{M}$, which has Jacobian $\det J = (\det M)^n$, after dropping the tilde the integral which determines the mean field partition function is given by
\begin{equation}
\label{eq: mfpartitionafter}
 \int dg \; e^{\gamma \sqrt{\det(M)} \left( \sqrt{\det \left(\frac{I + g}{2} \right)} - \sqrt{ \det g} \right)}.
\end{equation}
Thus we see that our mean field analysis results in a random matrix model which is invariant under $O(n)$ transformations. Because the matrix model has this symmetry group we can perform the diagonalization $ g = O^T \, \Lambda \, O$ where $\Lambda$ is the matrix of eigenvalues, all of which are by assumption positive. The Jacobian of this transformation is given by~\cite{Eynard:2000aa, Metha}
\begin{equation}
\label{eq: haarjacobian}
dg = dO \, \prod_{A} d \lambda_A \, \prod_{A<B} | \lambda_A - \lambda_B|.
\end{equation}
Thus, after integrating over the orthogonal group,  the integral appearing in the mean field theory partition function becomes
\begin{equation}
\int \prod_{A} d \lambda_A \; \prod_{A<B} | \lambda_A - \lambda_B| \, e^{-\gamma \, \sqrt{\det(M)} \, V(\lambda)}, 
\end{equation}
where the function $V(\lambda)$ is given by
\begin{equation}
V(\lambda) = \prod_A \sqrt{\lambda_A} - \prod_{A} \sqrt{\frac{1 + \lambda_A}{2}}.
\end{equation}
After a bit of work we obtain the (useful but sub-optimal) constraints
\begin{equation}
 \prod_{\lambda_A<1}  \sqrt{\lambda_A}    -    \prod_{\lambda_A>1}  \sqrt{\lambda_A} <  V(\lambda) < \prod_A \sqrt{\lambda_A}.
\end{equation}
Only for the trivial case $n = 1$ (a one-dimensional chain) is the function $V(\lambda)$ bounded from below; for $n\geq2$ there are directions in eigenvalue space where $V(\lambda)$ becomes arbitrarily negative. Thus the mean field truncation of the lattice action gives a random matrix model exhibiting pathology similar to that of the Einstein--Hilbert action. Further study of this model, using the techniques of random matrix theory, may shed light on how to deal with this feature.

\paragraph{Discussion:}

We have seen that with the particularly simple and elegant discrete action (\ref{eq: actiondef}) one can successfully encode a very large fraction of standard GR. The action is gauge fixed, with only rigid (global) rotations and parity inversions as symmetries,  and seems automatically to be in the de~Donder (nonlinear harmonic) gauge; this is a feature, rather than a problem --- the hyper-cubic lattice and the de~Donder gauge seem related at some deep level. Presumably there is some more general gauge-invariant action of which this is gauge-fixed version. We note also that this model does not exhibit the active diffeomorphism symmetry of the continuum theory and thus there are $n(n+1)/2$ dynamical degrees of freedom per lattice site; this is standard for discretisations of gravity \cite{diffeos}. 
Of course gauge fixing is not a problem \emph{per se}~\cite{fock}, since to make physical predictions one ultimately has to do so anyway. What is perhaps a little surprising is just how far one can get with such a simple and elegant discrete action. 

\paragraph{Acknowledegments:}
This research was supported by the Marsden Fund, and (MV) by a James Cook Research Fellowship, both administered by the Royal Society of New Zealand. 
KT acknowledges support by a Victoria University PhD scholarship.


\begin{thebibliography}{99} 
\bibitem{Regge:1961px} 
  T.~Regge,
  ``General Relativity Without Coordinates'',
  Nuovo Cim.\  {\bf 19}, 558 (1961).
  \bibitem{Williams:1996jb}
  R.~M.~Williams,
  ``Recent progress in Regge calculus'',
  Nucl.\ Phys.\ Proc.\ Suppl.\  {\bf 57}, 73-81 (1997).
  [gr-qc/9702006].
\bibitem{Sorkin:1975ah}
  R.~Sorkin,
  ``Time Evolution Problem in Regge Calculus'',
  Phys.\ Rev.\  D {\bf 12}, 385 (1975)
  [Erratum-ibid.\  D {\bf 23}, 565 (1981)].
\bibitem{Barrett:1994ks} 
  J.~W.~Barrett, M.~Galassi, W.~A.~Miller, R.~D.~Sorkin, P.~A.~Tuckey and R.~M.~Williams,
  ``A Paralellizable implicit evolution scheme for Regge calculus'',
  Int.\ J.\ Theor.\ Phys.\  {\bf 36}, 815 (1997)
  [gr-qc/9411008].
\bibitem{Rocek:1982fr} %
  M.~Ro\v{c}ek and R.~M.~Williams,
  ``Quantum Regge Calculus'',
  Phys.\ Lett.\  B {\bf 104}, 31 (1981).
\bibitem{Williams:1986hx}
  R.~M.~Williams,
  ``Quantum Regge Calculus in the Lorentzian Domain and its Hamiltonian Formulation'',
  Class.\ Quant.\ Grav.\  {\bf 3}, 853 (1986).
\bibitem{Hamber:1985gw}
  H.~W.~Hamber and R.~M.~Williams,
  ``Two-Dimensional Simplicial Quantum Gravity'',
  Nucl.\ Phys.\  B {\bf 267}, 482 (1986).
  \bibitem{Gross:1990fq}
  M.~Gross and H.~W.~Hamber,
  ``Critical properties of two-dimensional simplicial quantum gravity'',
  Nucl.\ Phys.\  B {\bf 364}, 703 (1991).
  \bibitem{David:1992jw}
  F.~David,
  ``Simplicial quantum gravity and random lattices'',
  arXiv:hep-th/9303127.
\bibitem{Ambjorn:1998xu}
  J.~Ambj\o{}rn and R.~Loll,
  ``Nonperturbative Lorentzian quantum gravity, causality and topology change'',
  Nucl.\ Phys.\  B {\bf 536}, 407 (1998)
  [arXiv:hep-th/9805108].
  \bibitem{Loll:2000my}
  R.~Loll,
  ``Discrete Lorentzian quantum gravity'',
  Nucl.\ Phys.\ Proc.\ Suppl.\  {\bf 94} (2001) 96
  [arXiv:hep-th/0011194].
  \bibitem{Ambjorn:2002gr}
  J.~Ambj\o{}rn, A.~Dasgupta, J.~Jurkiewicz, R.~Loll,
  ``A Lorentzian cure for Euclidean troubles'',
  Nucl.\ Phys.\ Proc.\ Suppl.\  {\bf 106 } (2002)  977-979.
  [hep-th/0201104].
  \bibitem{Ambjorn:2005db}
  J.~Ambj\o{}rn, J.~Jurkiewicz, R.~Loll,
  ``Spectral dimension of the universe'',
  Phys.\ Rev.\ Lett.\  {\bf 95 } (2005)  171301.
  [hep-th/0505113].
  \bibitem{Ambjorn:2005qt}
  J.~Ambj\o{}rn, J.~Jurkiewicz, and R.~Loll,
  ``Reconstructing the universe'',
  Phys.\ Rev.\  {\bf D72 } (2005)  064014.
  [hep-th/0505154].
  \bibitem{Ambjorn:2006jf}
  J.~Ambj\o{}rn, J.~Jurkiewicz, and R.~Loll,
  ``Quantum Gravity, or The Art of Building Spacetime'',
  arXiv:hep-th/0604212.
\bibitem{Tate:2011ct} 
  K.~Tate and M.~Visser,
  ``Fixed-Topology Lorentzian Triangulations: Quantum Regge Calculus in the Lorentzian Domain'',
  JHEP {\bf 1111}, 072 (2011)
  [arXiv:1108.4965 [gr-qc]].
\bibitem{Tate:2011rm} 
  K.~Tate and M.~Visser,
  ``Realizability of the Lorentzian $(n,1)$-Simplex'',
  JHEP {\bf 1201}, 028 (2012)
  [arXiv:1110.5694 [gr-qc]].
  \bibitem{Loll:1995mg} 
  R.~Loll,
  ``Nonperturbative solutions for lattice quantum gravity'',
  Nucl.\ Phys.\ B {\bf 444}, 619 (1995)
  [gr-qc/9502006].
\bibitem{Loll:1998aj}
  R.~Loll,
  ``Discrete approaches to quantum gravity in four-dimensions'',
  Living Rev.\ Rel.\  {\bf 1} (1998) 13
  [gr-qc/9805049].
\bibitem{exact}
H.~Stephani \emph{et al.},~\emph{Exact solutions of Einstein's field equations}, (Cambridge University Press, 2003).
\\
J.~B.~Griffiths and J.~Podolsk\'y, \emph{Exact spacetimes in Einstein's general relativity}, (Cambridge University Press, 2009).
\bibitem{acoustic}
C.~Barcel\'o, S.~Liberati and M.~Visser,
  ``Analogue gravity'',
  Living Rev.\ Rel.\  {\bf 8} (2005) 12
   [Living Rev.\ Rel.\  {\bf 4} (2011) 3]
  [gr-qc/0505065].
  \\
M.~Visser,
  ``Acoustic black holes: Horizons, ergospheres, and Hawking radiation'',
  Class.\ Quant.\ Grav.\  {\bf 15} (1998) 1767
  [gr-qc/9712010].
  \\
  M.~Visser,
  ``Acoustic propagation in fluids: An Unexpected example of Lorentzian geometry'',
  gr-qc/9311028.
\bibitem{fock}
V.~A.~Fock, \emph{The theory of space, time, and gravitation}, (Pergamon, London, 1959).
\bibitem{carmen}
 M.~Visser and C.~Molina-Par\'is,
  ``Acoustic geometry for general relativistic barotropic irrotational fluid flow'',
  New J.\ Phys.\  {\bf 12} (2010) 095014
  [arXiv:1001.1310 [gr-qc]].
\bibitem{petrov}
A.~N.~Petrov, 
 New harmonic coordinates for the Schwarzschild geometry and the field approach
Astronomical \& Astrophysical Transactions {\bf 1} (1992) 195--205.
\bibitem{massive-graviton}
 C.~de Rham and G.~Gabadadze,
  ``Generalization of the Fierz-Pauli Action'',
  Phys.\ Rev.\ D {\bf 82} (2010) 044020
  [arXiv:1007.0443 [hep-th]].
\\
  C.~de Rham, G.~Gabadadze, and A.~J.~Tolley,
  ``Resummation of Massive Gravity'',
  Phys.\ Rev.\ Lett.\  {\bf 106} (2011) 231101
  [arXiv:1011.1232 [hep-th]].
 \\
   S.~F.~Hassan and R.~A.~Rosen,
  ``On Non-Linear Actions for Massive Gravity'',
  JHEP {\bf 1107} (2011) 009
  [arXiv:1103.6055 [hep-th]].
\\
  S.~F.~Hassan and R.~A.~Rosen,
  ``Resolving the Ghost Problem in non-Linear Massive Gravity'',
  Phys.\ Rev.\ Lett.\  {\bf 108} (2012) 041101
  [arXiv:1106.3344 [hep-th]].
\\
V.~Baccetti, P.~Martin-Moruno, and M.~Visser,
  ``Massive gravity from bimetric gravity'',
  arXiv:1205.2158 [gr-qc].

\bibitem{Eynard:2000aa}
B.~Eynard,
``Lecture Notes on Random Matrices'',
Cours de Physique Th\'{e}orique de Saclay (2000).
[Saclay - T01/014, CRM-2708]
\bibitem{Metha}
M.~L.~Metha, \emph{Random matrices}, (Academic, New York,  2004).   ISBN-13: 978-0120884094.
\bibitem{diffeos}
  B.~Dittrich,
  ``Diffeomorphism symmetry in quantum gravity models'',
  arXiv:0810.3594 [gr-qc].
\\
 B.~Bahr and B.~Dittrich,
  ``(Broken) Gauge Symmetries and Constraints in Regge Calculus'',
  Class.\ Quant.\ Grav.\  {\bf 26}, 225011 (2009)
  [arXiv:0905.1670 [gr-qc]].
\end{thebibliography}
\end{document}